\documentclass[twocolumn]{aastex63}

\usepackage[T1]{fontenc}
\usepackage{ae,aecompl}

\usepackage{graphicx}	
\usepackage{amsmath}	
\usepackage{amssymb}	
\usepackage{url}

\def\arcdeg{\hbox{$^\circ$}}
\def\arcmin{\hbox{$^\prime$}}
\def\arcsec{\hbox{$^{\prime\prime}$}}
\def\flux{erg\,s$^{-1}$\,cm$^{-2}$}

\def\lum{erg s$^{-1}$}

\def\nustar{\textit{NuSTAR}}

\def\cgro{\textit{CGRO}}
\def\batse{BATSE}
\def\rxte{\textit{RXTE}}

\def\gro{GRO\,J2058+42}

\newcommand {\be}{\begin {equation}}
\newcommand {\ee}{\end {equation}}

\def\a{$^{\mbox{\small a}}$}
\def\b{$^{\mbox{\small b}}$}
\def\c{$^{\mbox{\small c}}$}
\def\d{$^{\mbox{\small d}}$}
\def\e{$^{\mbox{\small e}}$}

\received{XXX, 2019}
\revised{XXX, 2019}
\accepted{XXX}
\submitjournal{ApJL}

\shorttitle{Discovery of the cyclotron line in \gro}
\shortauthors{Molkov et al.}

\begin{document}

\title{Discovery of a pulse-phase-transient cyclotron line in the
X-ray pulsar \gro}

\correspondingauthor{Sergey Molkov}
\email{molkov@iki.rssi.ru}

\author{S. Molkov}
\affiliation{Space Research Institute, Russian Academy of Sciences,
Profsoyuznaya 84/32, 117997 Moscow, Russia}

\author{A. Lutovinov}
\affiliation{Space Research Institute,
 Russian Academy of Sciences,
Profsoyuznaya 84/32, 117997 Moscow, Russia}
\affiliation{Moscow Institute of Physics and Technology, Moscow region,
141701 Dolgoprudnyi, Russia}

\author{S. Tsygankov}
\affiliation{Department of Physics and Astronomy,
 FI-20014 University of Turku, Finland}
\affiliation{Space Research Institute,
 Russian Academy of Sciences,
Profsoyuznaya 84/32, 117997 Moscow, Russia}

\author{I. Mereminskiy}
\affiliation{Space Research Institute,
 Russian Academy of Sciences,
Profsoyuznaya 84/32, 117997 Moscow, Russia}

\author{A. Mushtukov}
\affiliation{Leiden Observatory, Leiden University,
 NL-2300RA Leiden, The Netherlands}
\affiliation{Space Research Institute,
 Russian Academy of Sciences,
Profsoyuznaya 84/32, 117997 Moscow, Russia}
\affiliation{Pulkovo Observatory, Russian Academy of Sciences,
Saint Petersburg 196140, Russia}

\begin{abstract}
We report the discovery of absorption features in the X-ray spectrum of the
transient X-ray pulsar \gro. The features are detected around $\sim10$,
$\sim20$ and $\sim30$ keV in both {\it NuSTAR} observations carried out
during the source type II outburst in spring 2019. The most intriguing
property is that the deficit of photons around these energies is registered
only in the narrow phase interval covering around 10\% of the pulsar spin
period. We interpret these absorption lines as a cyclotron resonant
scattering line (fundamental) and two higher harmonics. The measured energy
allow us to estimate the magnetic field strength of the neutron star as
$\sim10^{12}$\,G.

\end{abstract}




\section{Introduction} \label{sec:intro}

\gro\ is a slowly rotating ($P_{\rm spin}\simeq196$~s) transient X-ray pulsar
(XRP) discovered with the Burst and Transient Source Experiment (BATSE)
on board the {\it Compton Gamma-Ray Observatory} ({\it CGRO}) during a type II
(giant) outburst in 1995 September \citep[][]{1995IAUC.6238....1W}.  After
this outburst a dozen normal ones (type I) had been observed during the next two
years with  \cgro\ and the {\it Rossi X-Ray Timing Explorer} ({\it RXTE}). These
type I outbursts were spaced by about 110 day intervals, which was
interpreted as an orbital period of the system
\citep[][]{1996IAUC.6514....2W, 1997ApJS..113..367B}. At the same time
additional short and weak outbursts were detected by \batse\ halfway between
these type I outbursts \citep[][]{1998ApJ...499..820W}. Combining these
measurements with ones carried out with the  All-Sky  Monitor (ASM) on board
the \rxte\ observatory an alternative orbital period of $\sim 55$ days was
also considered \citep[][]{1998ApJ...499..820W,2005ApJ...622.1024W}.

The source localization accuracy ($30$\arcmin$\times60$\arcmin), obtained
with the \cgro\ \citep[][]{1995IAUC.6239....2G} and subsequently restricted
down to $4$\arcmin\ with \rxte\ \citep[][]{1996IAUC.6514....2W}, did not
allow us to make an immediate determination of the optical counterpart.
Only after the identification of \gro\ with the {\it Chandra} source
CXOU\,J205847.5+414637 and following observations in the optical band was the
normal companion reliably recognized as a Be star at a
distance of $9.0\pm1.3$ kpc \citep[][]{2005ApJ...622.1024W}.

Spectral properties of the source are poorly known. They were briefly
reported and discussed by \citet{2000AIPC..510..208W,2005ApJ...622.1024W}
using the \rxte/PCA and {\it Chandra} data and by \citet{2008ATel.1516....1K}
based on the {\it Swift X-Ray Telescope (XRT)} data. These authors
used an absorbed power law 
to describe the source spectrum in soft X-rays and the bremsstrahlung model
in a wider energy band.

In this Letter we perform a detailed spectral analysis of {\gro}  and report
the discovery of the cyclotron absorption line in its spectrum with the {\it
Nuclear Spectroscopic Telescope Array} (NuSTAR) observatory during the
type II outburst in the spring of 2019. For the first
time for accreting XRPs such a feature is robustly detected only in the
narrow range of pulse phases.

\begin{figure}
  \includegraphics[width=1.1\columnwidth]{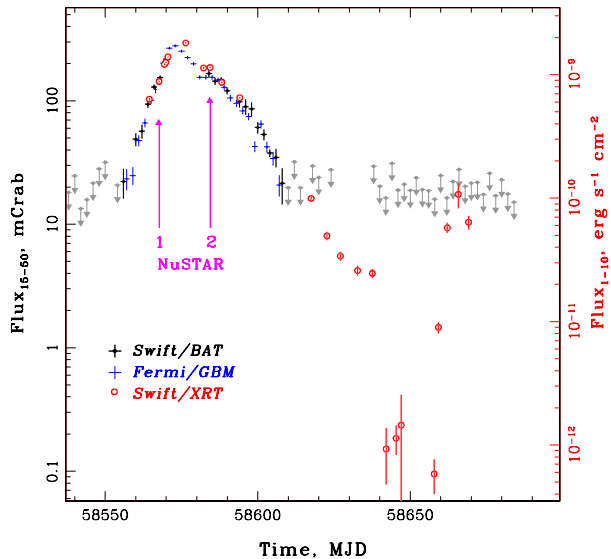}
  \caption{The {\it Swift}/BAT light curve (black crosses, 15-50 keV), {\it
  Fermi}/GBM pulsed emission (blue crosses, 12-50 keV), and {\it Swift}/XRT
  flux (red open circle, 1-10\,keV) measured from \gro\ during the 2019
  outburst. {\it Swift}/BAT data are in mCrab units (left axis), {\it
  Fermi}/GBM data are in keV\,cm$^{-2}$\,s$^{-1}$, and {\it Swift}/XRT data are
  in \flux\ (right axis). To trace the outburst morphology XRT and GBM curves
  are aligned with the BAT one at the moment of the second \nustar\ observation.
  Dates of two {\it NuSTAR} observations are marked with vertical magenta arrows.
}
\label{fig:outb_hist}
\end{figure}

\section{Observations and data reduction}

Since its discovery in 1995 during the giant outburst and subsequent two
years of activity, \gro\ remained in a quiet state until 2019. Only one weak
type I outburst was detected in 2008 May
\citep[][]{2008ATel.1516....1K}. The beginning of new type II outburst was
registered with the {\it Neil Gehrels Swift Observatory}
\citep[][]{2004ApJ...611.1005G} on 2019 March 22
\citep[][]{2019GCN.23985....1B} and later confirmed by the detection of the
pulsed emission \citep[][]{2019ATel12614....1M} with the Gamma-ray Burst
Monitor \citep[GBM; ][]{2009ApJ...702..791M} on board the {\it Fermi}
observatory.

This outburst lasted more than 100 days and was monitored by several
X-ray instruments. To trace the source light curve we used available data
from the {\it Swift}/BAT telescope \citep[][]{2013ApJS..209...14K} in the
15-50 keV energy band (Figure \ref{fig:outb_hist}).\footnote{\url{
https://swift.gsfc.nasa.gov/results/transients/weak/ \\ GROJ2058p42}} The
{\it Swift}/BAT data have a gap around the outburst maximum; therefore, to
better demonstrate an entire morphology of the outburst we used data of the
{\it
Fermi}/GBM\footnote{\url{https://gammaray.msfc.nasa.gov/gbm/science/pulsars/
\\ lightcurves/groj2058.html}} that were aligned with the BAT ones at the
moment of the second \nustar\ observation. Both light curves are in a good
agreement with each other (Figure \ref{fig:outb_hist}).

We also used data from the {\it Swift/}XRT
\citep[][]{2005SSRv..120..165B} to trace the evolution of the source flux in the
soft energy band. The fluxes measured with XRT in the 1-10 keV energy range
are shown in Figure \ref{fig:outb_hist} by red open circles. They were
calculated from the source spectra obtained with the online tools
\citep[][]{2009MNRAS.397.1177E}, provided by the UK Swift Science Data
Center.{\footnote{\url{http://www.swift.ac.uk/user_objects/}}}

The Nuclear Spectroscopic Telescope Array {\it NuSTAR} observatory consists
of two identical X-ray telescope modules, referred to as FPMA and FPMB
\citep{2013ApJ...770..103H}. It provides X-ray imaging, spectroscopy, and
timing in the energy range of 3-79\,keV with an angular resolution of
18\arcsec\ (FWHM) and spectral resolution of 400\,eV (FWHM) at 10\,keV. {\it
NuSTAR} performed two observations of \gro\ on 2019 March 25 (ObsID:
90501313002) and 2019 April 11 (ObsID: 90501313004) with the on-source
exposures of $\sim20$ and $\sim40$\,ks, respectively. Note that both
observations were carried out near the maximum of the outburst (see
Figure ~\ref{fig:outb_hist}, marked with the magenta arrows as ''1'' and ''2'').
The {\it NuSTAR} data were processed with the standard {\it  NuSTAR} Data
Analysis Software ({\sc nustardas}) v1.8.0 provided under {\sc heasoft} v6.25
with the {\sc caldb} version 20190513.

\begin{figure}
  \includegraphics[width=1.1\columnwidth]{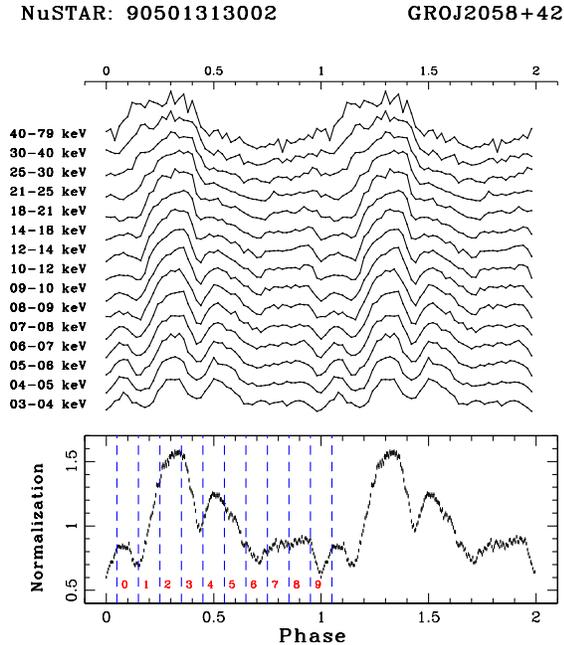}
\caption{Energy-resolved pulse profiles of \gro\ obtained with {\it NuSTAR}
in the first observation. In the bottom panel an averaged pulse profile is
shown. Vertical lines demonstrate phase bins selected for spectral analysis.}
\label{fig:fold_crv}
\end{figure}

In the following spectral analysis we used the $4-79$ keV energy band. An
increase of the lower threshold energy from the standard 3 to 4 keV is due to
both observations being made during the solar activity periods. It could
affect the correctness of the background estimation with standard routines
below 4 keV, where the background could dominated by a few lines
and the $\sim1$\,keV thermal plasma component, probably connected with reflected
solar X-rays \citep[][]{2014ApJ...792...48W}.

All obtained spectra were grouped to have at least 25 counts per bin using
the {\sc grppha} tool. The final data analysis (timing and spectral) was
performed with the {\sc heasoft 6.25} software package. All uncertainties are
quoted at the {\bf 90\%} confidence level, if not stated otherwise.

\section{Results}

We performed a complete timing and spectral (including pulse-phase-resolved)
analysis for both {\it NuSTAR} observations. Resulting spectra and pulse
profiles of the source are very similar each other and for briefness we
present most of following figures only for the first observation.

\subsection{Energy-resolved pulse profile}

Orbital ephemerides for \gro\ are unknown; therefore, the pulsating signal was
searched only in barycentered light curves. Pulsations were clearly detected
with the high significance at periods of $195.240(2)$ and $194.149(1)$~s for
the first and second {\it NuSTAR} observations, respectively. These values
were used in the subsequent analysis to fold light curves and for the
pulse-phase-resolved spectroscopy.

Figure~\ref{fig:fold_crv} presents energy-resolved pulse profiles of the
source obtained in the first {\it NuSTAR} observation. We attributed the
phase 0 to the minimum of the folded light curve in the full instrument
energy band. The pulse profile is clearly evolving with the energy.

At the few to about 10 keV energy range, the profile shows three distinct
peaks at phases 0.1, 0.3, and 0.5. As the energy increases the two
"outer" peaks disappear and the central peak eventually expands,
while its minimum shifts to the phase $\sim0.7$.

The pulsed fraction gradually increases with the energy
from $\sim 40\%$ at $3-5$ keV to $\sim 60\%$ at
$50-70$ keV, which is observed for the majority of bright XRPs \citep[see,
e.g., ][]{2009AstL...35..433L}.

\subsection{Phase-averaged spectrum}

\begin{figure}
  \includegraphics[width=1.1\columnwidth]{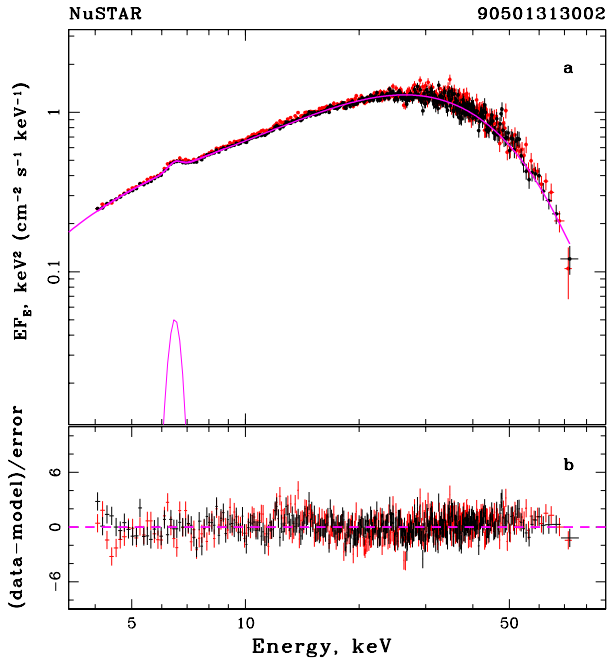}
  \caption{(a) Broadband energy spectrum of \gro\ obtained in the first
  {\it NuSTAR} observation. Black and red crosses correspond to FPMA and
  FPMB modules. Blue solid line represents the best-fit model (see details
  in the text). (b) Residuals from the best-fit model.}
\label{fig:mean_spec}
\end{figure}

The spectrum of \gro\ has a typical shape for accreting XRPs \citep[see,
e.g.,][]{1989PASJ...41....1N, 2005AstL...31..729F} and demonstrates an
exponential cutoff at high energies (Figure \ref{fig:mean_spec}(a)), that, e.g.,
can be explained in terms of the Comptonization processes in hot emission
regions \citep[see, e.g., ][]{1980A&A....86..121S, 1985ApJ...299..138M}.
Therefore, at the first stage it was approximated with several commonly used
models: a power law with an exponential cutoff (\texttt{cutoffpl} in the {\sc
xspec} package), a power law with a high-energy cutoff
(\texttt{powerlaw*highcut}), and a thermal Comptonization (\texttt{comptt}).
To take into account the uncertainty in the calibrations of two modules of
{\it NuSTAR} the cross-calibration constant $C$ between them was included in
all spectral fits. It was found that the Comptonization model
\citep[][]{1994ApJ...434..570T} with an inclusion of the iron emission line at 6.4
keV in the form of the Gaussian profile describes the \gro\ spectrum
significantly better than other models ($\chi^2=2255$ for 2117 degrees of
freedom (dof) in a comparison with 3730 (2120 dof) and 3911 (2119 dof)
for the first two models). Results of the approximation of the
source spectrum obtained in the first {\it NuSTAR} observation with this
model are show in Figure \ref{fig:mean_spec}a. Best-fit parameters are as
follows: the seed photons temperature $kT_{0}=1.55\pm0.15$\,keV, the plasma
temperature $kT=10.25\pm0.04$\,keV, the plasma optical depth $\tau=5.02\pm0.03$, the
iron line energy $E_{\rm Fe}=6.48\pm0.03$\,keV, the iron line width
$\sigma_{\rm Fe}=0.24\pm0.03$\,keV, its equivalent width $EW_{\rm
Fe}=70\pm9$\,eV, the total flux in the 4-79 keV energy range
$F_{4-79}\simeq3.6\times10^{-9}$\,\flux. From the bottom panel 
(Figure \ref{fig:mean_spec}b) it is seen that this model describes the spectrum 
adequately, and no obvious additional components are required.  Note that
it is difficult to compare directly the results of our measurements with ones 
obtained earlier \citep{2005ApJ...622.1024W,2008ATel.1516....1K}, as the source 
was observed in different intensity states in different energy bands. Nevertheless,
if we are restricted to only soft X-rays ($<10$\,keV) and used the
power-law model 
its parameters will be comparable with previously reported results.

The second {\it NuSTAR} observation was performed at the decaying part of the
outburst (Figure \ref{fig:outb_hist}) at a similar source intensity to the
first one, $F_{4-79}\simeq4.3\times10^{-9}$\,\flux. Spectral parameters
measured in this observation agree well with those reported above, and
again no additional components are required to describe the source spectrum.

\begin{figure}
\vbox{
  \includegraphics[width=1.1\columnwidth]{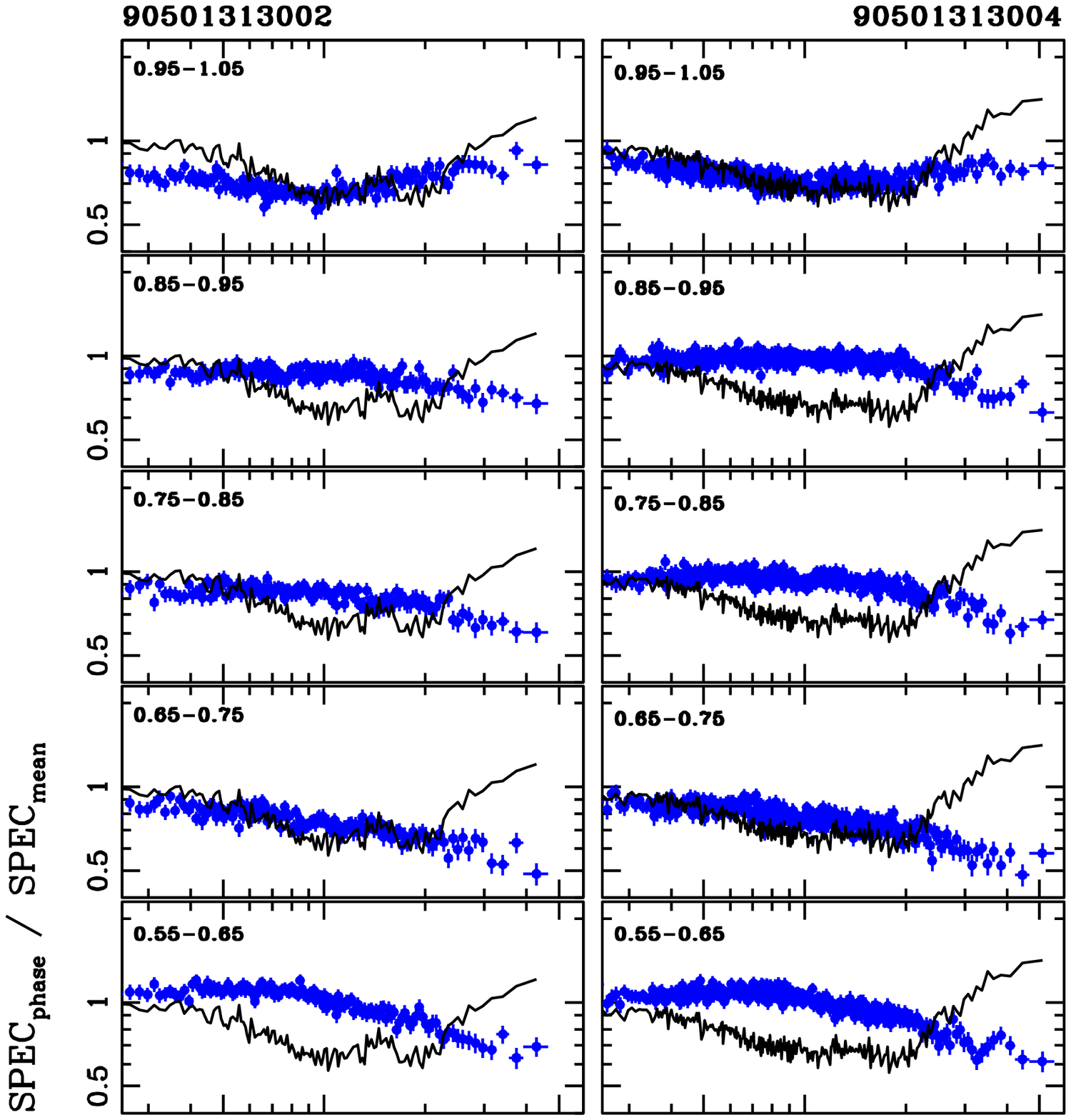}

  \vspace{-1.38cm}

\includegraphics[width=1.1\columnwidth]{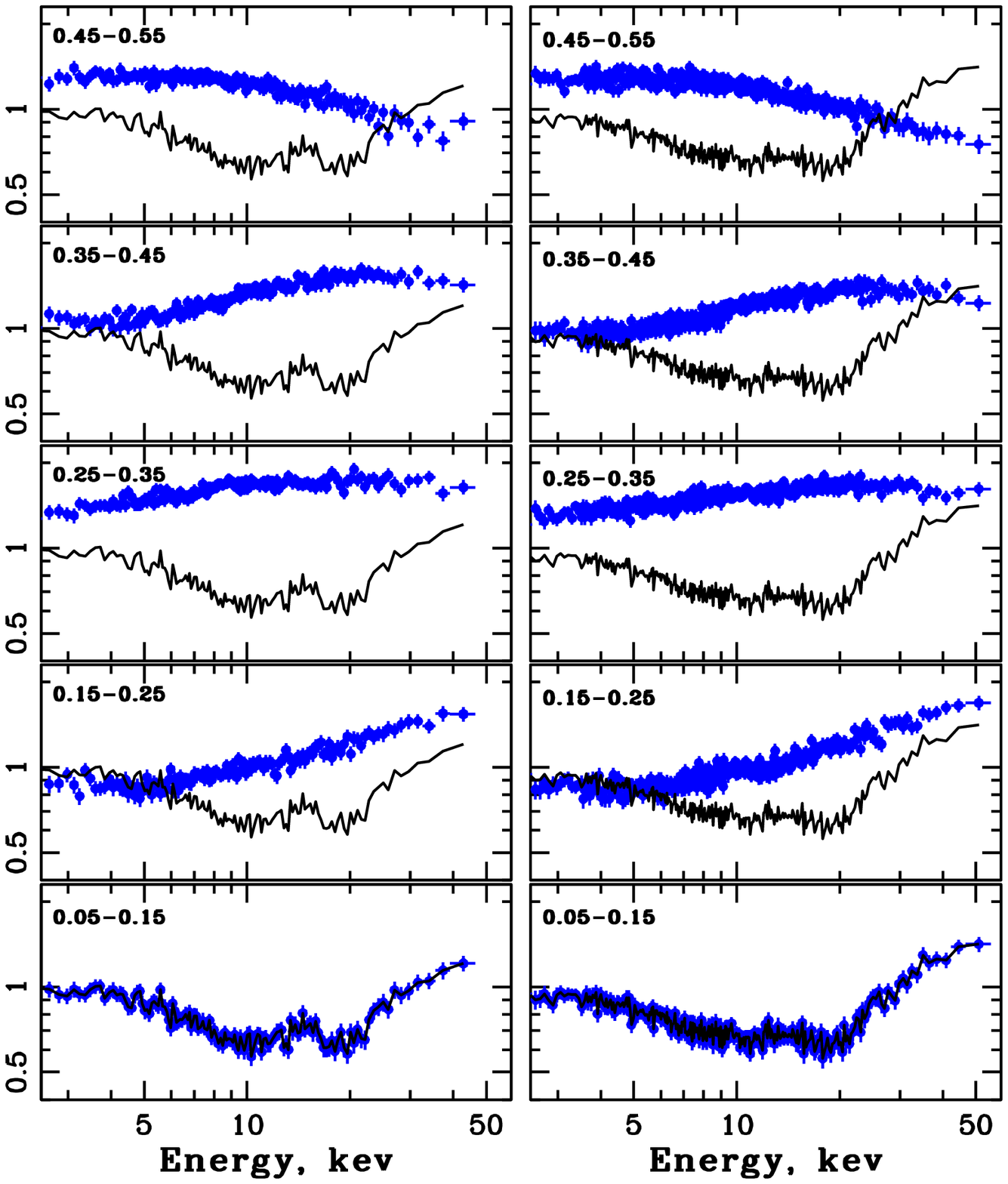}
}
\caption{ Ratio of the source spectra, measured at a given phase, to the averaged
one (blue points) for two {\it NuSTAR}
observations of \gro\ (left panels correspond to the observation 1, right
ones - to the observation 2). For comparison, the
ratio at the phase 0.05-0.15 is shown in each panel with thin black lines.}

\label{fig:ph_spc_evol}
\end{figure}

\subsection{Pulse-phase-resolved spectroscopy}

It is well established that spectra of XRPs are significantly variable with
the pulse phase. Parameters of the cyclotron resonant scattering features
(CRSFs), if they are present in the spectra, also change \citep[see,
e.g.,][and references
therein]{2000ApJ...530..429B,2004A&A...427..975K,2004AIPC..714..323H,2015MNRAS.448.2175L}.
Therefore, the pulse-phase-resolved spectroscopy can be considered as a tool
for the diagnosis of the geometry of the emission regions in the vicinity of
the neutron star and its magnetic field structure. To trace an evolution of
the \gro\ spectrum with the pulse phase we used the ratio of each phase's
spectrum to the pulsed-averaged one. It is important to note that the result
of such an approach does not depend on the specific spectral model.

Results of the analysis are shown in Figure \ref{fig:ph_spc_evol}. It is
clearly seen that the source spectrum varies significantly with the pulse
phase, primarily demonstrating an evolution of its hardness. In particular,
the spectrum is hardest at the phases of $0.95-0.25$ where the small
interpeak is observed (see Figure \ref{fig:fold_crv}). The spectra become
gradually softer to the maximum of the first peak and to the second peak
(phases $0.45-0.55$) and returning to the hard state further. It is important
to note that such a behaviour is identical for both observations
(Figure \ref{fig:ph_spc_evol}).

\begin{figure}
  \includegraphics[width=1.1\columnwidth]{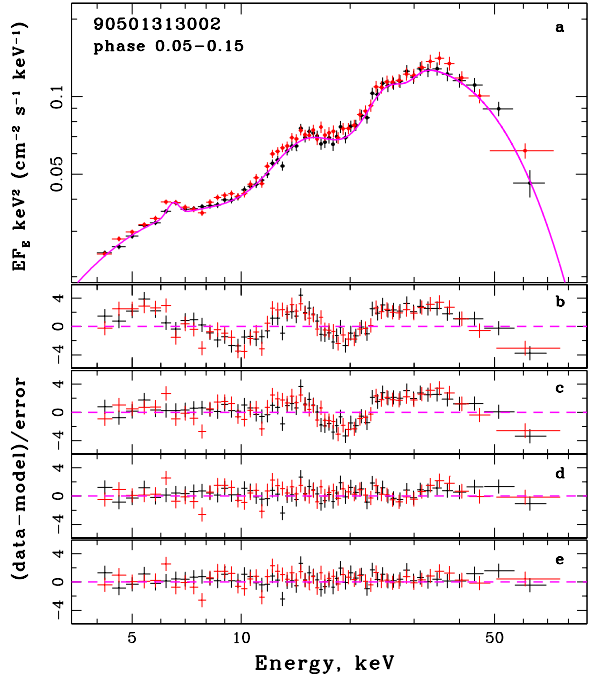}
\caption{Energy spectrum of \gro\ at the pulse phases 0.05-0.15 for the
first {\it NuSTAR} observation. The data from both the FPMA and FPMB modules are
shown by black and red points, respectively. Residuals in the bottom panels
demonstrate the quality of fits with four different models (see the text for
details).} \label{fig:cycl_spec}
\end{figure}

The most exceptional spectra are registered at phases of $0.05-0.15$ in both
observations, where an obvious deficit of photons around $\sim10$ and
$\sim20$ keV is observed (Figure \ref{fig:ph_spc_evol} and
\ref{fig:cycl_spec}a). To quantify these features the spectrum from the first
{\it NuSTAR} observation was fitted with different models. First of all we
used the simplest model adequately describing the averaged spectrum (\texttt{comptt+gaus}),
resulting in an unacceptable fit with $\chi^2=1449.6$ for 1127 dof and
obvious residuals around $\sim10$ and $\sim20$ keV
(Figure \ref{fig:cycl_spec}(b)). The successive inclusion of additional CRSF
components in the form of the {\texttt{gabs}} model significantly improves
the fit quality: up to $\chi^2=1310.1$ (1124 dof) with the line around
$\sim10$ keV (Figure \ref{fig:cycl_spec}(c)) and up to $\chi^2=1103.0$ (1121
dof) with two lines at $\sim10$ and $\sim20$ keV
(Figure \ref{fig:cycl_spec}(d)). Moreover, there is a marginal hint for the
presence of an additional weak absorption feature around $\sim30$ keV
(Figure \ref{fig:cycl_spec}e, fit quality is $\chi^2=1094.6$ for 1118
dof).

Similar absorption features at the same energies also are registered also in the source
spectrum reconstructed for the same pulse phases in the second {\it NuSTAR}
observation, but in this case an additional third absorption line at $\sim30$
keV improves the fit more significantly, from $\chi^2=1584$ (1531 dof) to
$\chi^2=1547.8$ (1528 dof).

We interpreted these features as a cyclotron absorption line at $\sim10$ keV
with two higher harmonics, with parameters that can be summarized as in Table 1.

\medskip
\begin{center}
\begin{tabular}{lccc}
\hline
ObsID & $E_{\rm c}$, keV & $\sigma_{\rm c}$, keV & $\tau_{\rm c}$ \\
\hline
90501313002 & $10.00^{+0.27}_{-0.61}$ & $2.63^{+0.99}_{-0.38}$ & $0.34^{+0.51}_{-0.10}$ \\
            & $19.47^{+0.22}_{-0.52}$ & $3.23^{+0.39}_{-0.44}$ & $0.42^{+0.14}_{-0.08}$ \\
            & $28.23^{+1.00}_{-2.43}$ & $2.11^{+2.75}_{-0.87}$ & $0.12^{+0.21}_{-0.07}$ \\
90501313004 & $10.91^{+0.62}_{-0.48}$ & $3.14^{+2.18}_{-0.61}$ & $0.24^{+0.43}_{-0.08}$ \\
            & $19.40^{+0.42}_{-0.44}$ & $3.33^{+0.52}_{-0.54}$ & $0.49^{+0.09}_{-0.14}$ \\
            & $28.31^{+0.97}_{-1.93}$ & $3.40^{+1.70}_{-0.90}$ & $0.18^{+0.16}_{-0.07}$ \\
\hline
\end{tabular}
\end{center}
\medskip

\noindent
where $E_{\rm c}$, $\sigma_{\rm c}$ and $\tau_{\rm c}$ are the energy, width,
and optical depth of the cyclotron line and its higher harmonics.

To estimate the detection significance for each absorption feature we
performed three 10$^4$ Monte Carlo simulations of
the source spectra, successively adding the first, second, and
third {\texttt gabs} components.
We found that for the first observation the probabilities of
chance occurrence of 10, 20, and 30 keV features
are $<10^{-4}$,  $<10^{-4}$, and  $0.1145$, respectively.
For the second observation corresponding probabilities
are $<10^{-4}$,  $<10^{-4}$, and $10^{-4}$. Taking into account that the lines
are registered independently in two observations at the same
energies, the joint probabilities that they originate by chance
are significantly lower.

We made a detailed search for any absorption features in spectra at other
pulse phases, but all of them can be well described with the simple model,
used for averaged spectra, and no additional absorption lines are required.
To increase statistics we also constructed the spectrum averaged over all
phases with the exception of data at phases $0.05-0.15$, and again we found no
indication of the presence of the absorption lines in this spectrum.

\section{Discussion and conclusions}

Here we present the first robust detection of the CRSF localized in a very
narrow range of the spin phases of \gro\ and covering only $\sim$10\% of the
entire spin period. Previous evidence of a similar transient CRSF,
detectable in a small fraction of the pulsar rotation, was revealed in
spectra of several isolated neutron stars  \citep[see e.g.
][]{2015ApJ...807L..20B}. However, in the classical XRPs only a hint for the
marginal detection of such a feature was reported for EXO\,2030+375 based on
the {\it INTEGRAL} data \citep{2008A&A...491..833K}.

To explain the peculiar spectral properties of \gro\ one can consider a
geometrical configuration of the system.

High mass accretion rates onto the surface of neutron stars in XRPs result in
the appearance of accretion columns confined by a strong magnetic field of
the neutron star and supported by a high internal radiation pressure
\citep{1976MNRAS.175..395B,1981A&A....93..255W,2015MNRAS.454.2539M}. Thus,
the cyclotron line can originate from the accretion column
\citep{2014ApJ...781...30N,2015ApJ...807..164N,2014A&A...564L...8S} or it can
be a result of the reflection of X-rays from the atmosphere of the neutron
star \citep{2013ApJ...777..115P,2015MNRAS.448.2175L}. Due to a large gradient
of the $B$-field strength over the visible column height (see, e.g.,
\citealt{2015ApJ...807..164N}) or latitudes on the stellar surface, the
scattering feature can vanish from the observed energy spectra. However, a
situation where the accretion column is partially eclipsed by the neutron
star at certain phases of the pulse and the observer sees only a fraction of
the accretion column is possible \citep{2018MNRAS.474.5425M}. In this case,
the dispersion of the magnetic field strength over the visible part of the
column is relatively small, and the cyclotron line can appear at some
phases of
pulsations as it is observed in \gro.

It is also necessary to note that the visibility of both the neutron star
surface and accretion
column is strongly affected by the effects of gravitational light bending
(see e.g. \citealt{1988ApJ...325..207R,2001ApJ...563..289K,2018MNRAS.474.5425M}).
Remarkably, the column located on the opposite side of the neutron star
tends to provide
the majority of the observed X-ray energy flux due to effects of the
gravitational lensing.
In general, the pulse profile and spectrum of XRPs at supercritical
luminosities are determined by a large number of factors
including the accretion column height, compactness of the central
object, angular distribution of initial X-ray photons at the
stellar surface, edges of accretion column, etc. All of these factors
have to be included in an accurate theoretical model.

Considering the $E_c\simeq10$ keV feature as a fundamental energy of the
cyclotron absorption line, the magnetic field in the emission region can be
estimated as $B\sim10^{12}$\,G.

Another independent way to estimate independently the magnetic field of the
neutron star is to consider its quiescent luminosity and long-term flux
behavior. In particular, it was shown that the transition to the so-called
propeller regime \citep{1975A&A....39..185I}, when the rotating magnetosphere
centrifugally inhibits the accretion process, can be used to determine a
dipole component of the magnetic field of the neutron star
\citep{2016A&A...593A..16T,2016MNRAS.457.1101T,2017ApJ...834..209L}. After
the transition to the propeller regime the source spectrum becomes much
softer with the blackbody temperature of $\sim 0.5$~keV and quiescent
luminosity of $\sim 10^{33}$~erg~s$^{-1}$
\citep{2016MNRAS.463L..46W,2016A&A...593A..16T,2017MNRAS.470..126T}. However,
as it was shown later by \citet{2017A&A...608A..17T}, a transition to the
propeller regime is possible only for relatively fast spinning XRPs ($P_{\rm
spin}\lesssim10$~s). In the slowly rotating pulsars (like \gro) the accretion
disk switches to the ``cold'' low-ionization state maintaining a stable mass
accretion rate around $10^{14-15}$~g~s$^{-1}$. This rate depends on the inner
radius of the disk \citep{2017A&A...608A..17T} and therefore can be utilized
to estimate the magnetic field in XRPs
\citep{2019A&A...621A.134T,2019A&A...622A.198N}. Note that an analogous
physical mechanism was proposed earlier for cataclysmic variables
\citep[see, e.g.,][]{2001NewAR..45..449L}.

As can be seen from Figure \ref{fig:outb_hist} \gro\ switched to the quiescent
state around MJD\,58640. This state is characterized by a stable low-level
flux around $10^{-12}$\,\flux, that corresponds to the luminosity around
$10^{34}$\,\lum, assuming a distance to the source of 9 kpc. It is worth noting
that a serendipitous {\it Chandra} detection of the source on 2004 February 24
resulted in the same flux, pointing to the quiescent nature of this emission.
Important information about the emission mechanism can be derived from the
spectral analysis in the quiescent state; however, available data do not allow
us to robustly discriminate between the soft blackbody-like and hard
accretion-like spectral models. However, as discussed above, this luminosity
is too high for the propeller regime and can be interpreted as a stable
accretion from the cold disk. In this case we can use  Eq.\,(7) from
\citet{2017A&A...608A..17T} to estimate the magnetic field in the neutron star in
\gro. Assuming a standard mass and radius of the neutron star and the source
distance of 9 kpc, we get a magnetic field strength around
$(1-2)\times10^{12}$\,G, which is in very good agreement with the value derived
from the cyclotron energy.

\section*{Acknowledgements}

We thank the {\it NuSTAR} and {\it Swift}/XRT teams for organizing
prompt observations. This work was financially supported by the Russian
Science Foundation (grant 19-12-00423).




\bibliography{gro}
\bibliographystyle{aasjournal}
\label{lastpage}
\end{document}